\pdfoutput=1
\documentclass[12pt, letterpaper]{article}
\usepackage{graphicx}
\usepackage{xcolor}
\usepackage[font=small,labelfont=bf,justification=justified]{caption}
\usepackage{gensymb}
\usepackage[toc,page]{appendix}
\usepackage[utf8]{inputenc}
\usepackage{amsmath}
\usepackage{dsfont}
\usepackage{wrapfig}
\usepackage{sectsty}
\usepackage{titlesec}
\usepackage{mathtools}
\usepackage{siunitx}
\usepackage{amssymb}
\usepackage{comment}
\DeclarePairedDelimiterXPP\BigOSI[2]%
  {\mathcal{O}}{(}{)}{}%
  {\SI{#1}{#2}}
\sectionfont{\fontsize{14}{15}\selectfont}
\usepackage[hyperfootnotes=false]{hyperref}
\hypersetup{
    citecolor=Violet,   
    filecolor=Red,      
    linkcolor=Blue,      
    urlcolor=cyan       
}
\usepackage[export]{adjustbox}
\usepackage[
backend=biber,
style=authoryear,
maxnames=1 % Limiting to display only the first author
]{biblatex}
\usepackage[a4paper, total={6.5in, 10in}]{geometry}
\title{Linking Mixing Interface Deformation to Concentration Gradients in Porous Media}
\author{Saif Farhat, Guillem Sole-Mari,  Diogo Bolster}
\date{\today}

\begin{document}
\maketitle
\begin{abstract}
We study the pore-scale transport of a conservative scalar forming an advancing mixing front, which can be re-interpreted to predict instantaneous mixing-limited bimolecular reactions. We investigate this using a set of two-dimensional, high-resolution numerical simulations within a poly-disperse granular porous medium, covering a wide range of Peclet numbers. The aim is to show and exploit the direct link between pore-scale concentration gradients and mixing interface (midpoint concentration isocontour). 
%We show that the length of the mixing interface at any time is the integral of the pore-scale gradient magnitudes over the entire mixing area/volume. 
We believe that such a perspective provides a complementary new lens for better understanding mixing and spreading in porous media. We  develop and validate a model that quantifies the temporal elongation of the mixing interface and the upscaled reaction kinetics in mixing-limited systems accounting for pore-scale concentration fluctuations. Contrary to the classical belief that, given sufficient time, pore-scale fluctuations would eventually be washed out, we show that for $Pe>1$ advection generates pore-scale concentration fluctuations more rapidly than they can be fully dissipated. For such P\'eclet numbers, once incomplete mixing is established, it will persist indefinitely.
\end{abstract}

\section{Introduction}
Solute transport, mixing, and reactions in porous media are fundamental processes for many natural and industrial applications (\cite{dentz2011mixing,rolle2019mixing,valocchi2019,dentz2023mixing}). Many reactions are driven by mixing and are considered mixing-limited when reaction time scales are shorter than mixing times (\cite{chiogna2012mixing,engdahl2014predicting,villermaux2019mixing}). This is often the case for many relevant systems and processes, including groundwater contaminant transport and degradation (\cite{kang2019potential}), mineral dissolution and precipitation (\cite{al2019pore,cil2017solute}), and CO$_2$ sequestration (\cite{macminn2012spreading}). To date, classical Darcy-scale and Advection-Dispersion reactive transport models have been the conventional framework to describe these processes in porous media applications (\cite{dentz2011mixing,valocchi2019}). However, in their traditional forms, these models do not account for naturally occurring fluctuations at the pore scale (i.e. incomplete mixing) and the influence these have on mixing-limited reactions (\cite{raje2000experimental, gramling2002reactive}). Accounting for pore-scale concentration fluctuations is crucial to predicting large-scale behaviors, and disregarding them is a significant cause for discrepancies between observations and model predictions in reactive transport scenarios (\cite{ willingham2008evaluation,battiato2011applicability,anna2014mixing,ding2017elimination,engdahl2017lagrangian,farhat2024evolution}).

In this context, the deformation of mixing interfaces and the pore-scale concentration fluctuations are widely recognized as key factors in determining global reaction kinetics (\cite{ranz1979applications,ottino1989kinematics,le2010non,de2012time,de2014filamentary}). Mixing interfaces within scalar fields are subjected to simultaneous stretching and shrinking mechanisms (\cite{hallack20243dporescalemixinginterface}) arising from pore-scale velocity variations and diffusion% (before and after coalescence)
, respectively, leading to the formation of elongated lamellar structures. %, with concentration fluctuations occurring at scales defined by the Batchelor scale (\cite{souzy2018mixing, villermaux2019mixing})
In some cases, stretching is enhanced by features such as chaotic advection and the presence of stagnation points. While the former can only occur in systems with sufficient degrees of freedom such as three-dimensional porous media (\cite{lester2016chaotic,heyman2020stretching}), stagnation points are common to both two and three dimensional ones. 
Trapping of fluid elements near the stagnation points induces significant folding and neighboring elements are swept downstream. Also,  in flows with spatially heterogeneous  shear rates, the elongation of the material line can follow a range of sub-exponential growth rates (\cite{dentz2016coupled}). Just like mixing interfaces, pore-scale concentration gradients are controlled by the competing effects of fluid stretching and diffusion, which enhance and dissipate them, respectively (\cite{ranz1979applications, villermaux2003mixing, duplat2008mixing, le2015lamellar}). 

A mixing process between two solutions may be modeled as essentially governed by $i$) a mixing front or interface through which the diffusive flux of solutes occurs, and $ii$) a normal concentration gradient that determines the rate of said diffusive flux (\cite{de2014filamentary}). In the context of mixing-limited reactions, the diffusive flux rate across the mixing interface controls the rate at which reactants come into contact with each other, and therefore affects the effective reaction rate. As expounded earlier, both the mixing interface length/area and the typical concentration gradient across it are dynamic quantities that can change over time.
Many studies have highlighted the importance of the coupling between the structure of the velocity field fluctuations and the resulting scalar field gradients (\cite{kraichnan1974convection,kraichnan1994anomalous,balkovsky1999universal,le2015lamellar,le2017scalar}). However, for applications concerned with reactive transport in porous media, the direct link between a scalar field's mixing interface deformation and concentration gradients is, to date, less clear.  It has been observed that under Poiseuille flow, or in fully saturated porous media, the mixing interface initially undergoes elongation until an equilibrium between stretching and shrinking is achieved. At this point, the interface neither elongates nor shrinks (\cite{hallack20243dporescalemixinginterface, jimenez2015pore}). In a way, this observation %contradicts 
challenges the classical view of Taylor dispersion, which states that any transverse variation in the concentration field under a fully developed Poiseuille flow ultimately degenerates to a uniform concentration (\cite{taylor1953dispersion}).
In this paper we aim to demonstrate and exploit the direct link between the temporal evolution of mixing interface deformation and concentration gradients in mixing fronts. By doing so, we seek to (1) offer an alternative view of of the Taylor-Aris dispersion process and (2) analytically quantify the product mass of a bimolecular mixing-limited reaction in porous media, while accounting for incomplete mixing.

\section{Simulations}
\begin{figure}
    \centering
    \includegraphics[width=\textwidth]{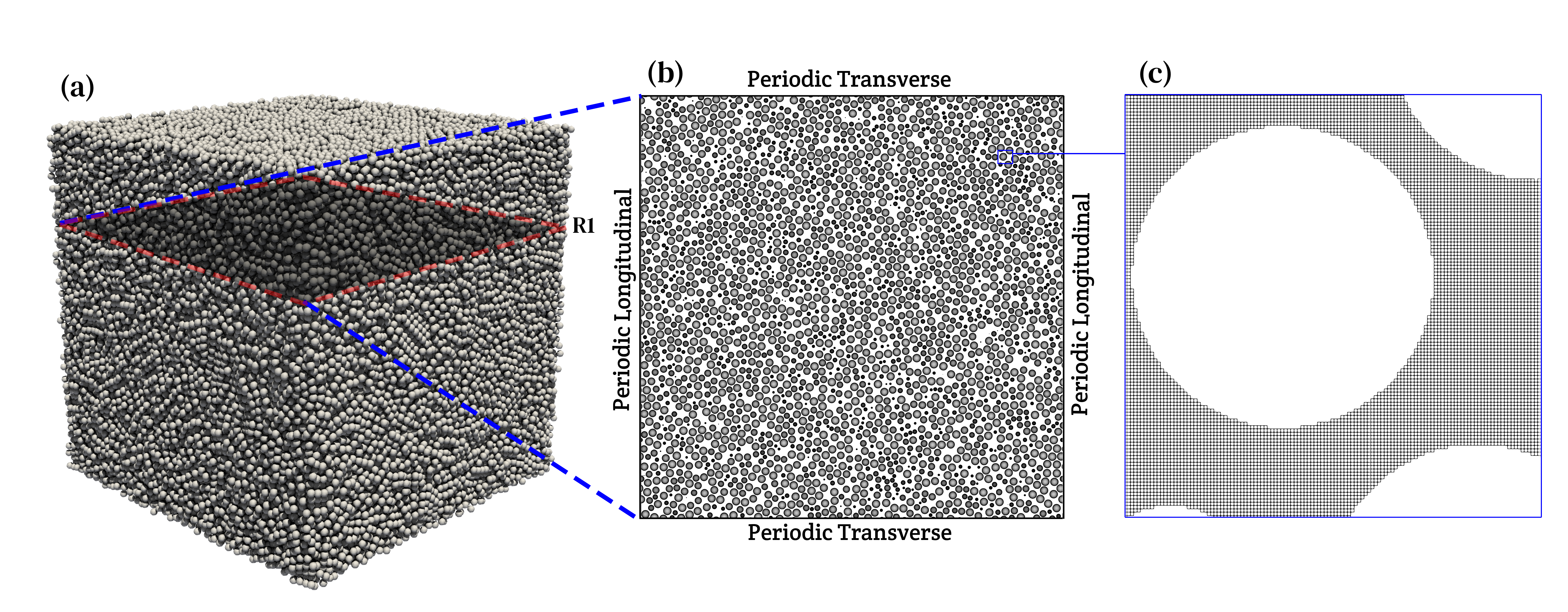}
    \caption{(a) Periodic cubic random packing of mono-dispersed spherical grains (50$d_0$, 50$d_0$, 50$d_0$); 
    (b) Slice taken from the three-dimensional column representing the first realization (R1); 
    (c) Finite volume structured mesh that was used in the simulations ($\Delta = d_0/100$).}
    \label{fig:Column}
\end{figure}

We generate multiple random realizations of two-dimensional porous media, each with identical statistical properties, to ensure representative simulation results. This is achieved by extracting two dimensional slices from a three-dimensional random porous medium. The latter consists of a random close packing of spheres with a mono-disperse diameter $d_0=2\mathrm{mm}$, generated using the algorithm developed by \cite{skoge2006packing}, as illustrated in Figure \ref{fig:Column} (a). The medium is periodic in all directions to prevent boundary effects. Five two-dimensional domains are extracted; each has dimensions of 50$d_0$ x 50$d_0$ and a porosity ($n$) of approximately $0.48$. The two-dimensional grains resulting from the slicing are poly-disperse with an average diameter of 1.5 mm. Figure \ref{fig:Column} (b) shows the first realization (R1) with others taken as slices at other heights.

The simulations are performed using multiple modules from OpenFOAM v10 (\cite{weller1998tensorial}). To generate the pore space mesh, a background structured hexahedral mesh is initially created using the {\fontfamily{qcr}\selectfont blockMesh} utility. Subsequently, the {\fontfamily{qcr}\selectfont castellatedMesh} feature within the {\fontfamily{qcr}\selectfont snappyHexMesh} utility is used to remove cells occupied by circular grains. The mesh has a grid size of $\Delta$ = $d_0$/100. Figure \ref{fig:Column} (c) shows a zoomed-in example region of the mesh.

To obtain the steady-state velocity field, the {\fontfamily{qcr}\selectfont simpleFOAM} algorithm was used to numerically solve the single-phase, steady-state, incompressible Navier-Stokes (NS) equations,
\begin{equation}
\label{div}
\nabla \cdot \pmb{u}=0 \: ,
\end{equation}
\begin{equation}
\label{momentum}
\pmb{u} \cdot \nabla \pmb{u}=-\frac{1}{\rho}\nabla p + \nu \nabla^2 \pmb{u}+\frac{1}{\rho} \pmb{f},
\end{equation}
\noindent
where {$\pmb{u}$} and {$\nabla p$} are the velocity field and the pressure gradient, respectively; $\rho$ is fluid density and {$\nu$} is kinematic viscosity. 
To drive the flow we apply a body force $\pmb{f}=[f,0]$ to maintain a specific, volume-averaged, pore flow velocity ($\bar{u}$).  In order to achieve conditions representative of water flowing in natural porous media, which is a process typically dominated by viscous effects (i.e stokes flow), we set the value of the longitudinal body force $f$, as well as those of $\rho$ and $\nu$, to attain a relatively low Reynolds number $Re=\bar{u}d_0/\nu=0.15$.
Periodic boundary conditions are imposed on both the longitudinal and transverse walls as shown in Figure \ref{fig:Column} (b). No-slip boundary conditions are enforced at the contact between the fluid and the solid grains. 

\begin{figure}
    \centering
    \includegraphics[trim=0 500 0 0, clip, width=\textwidth]{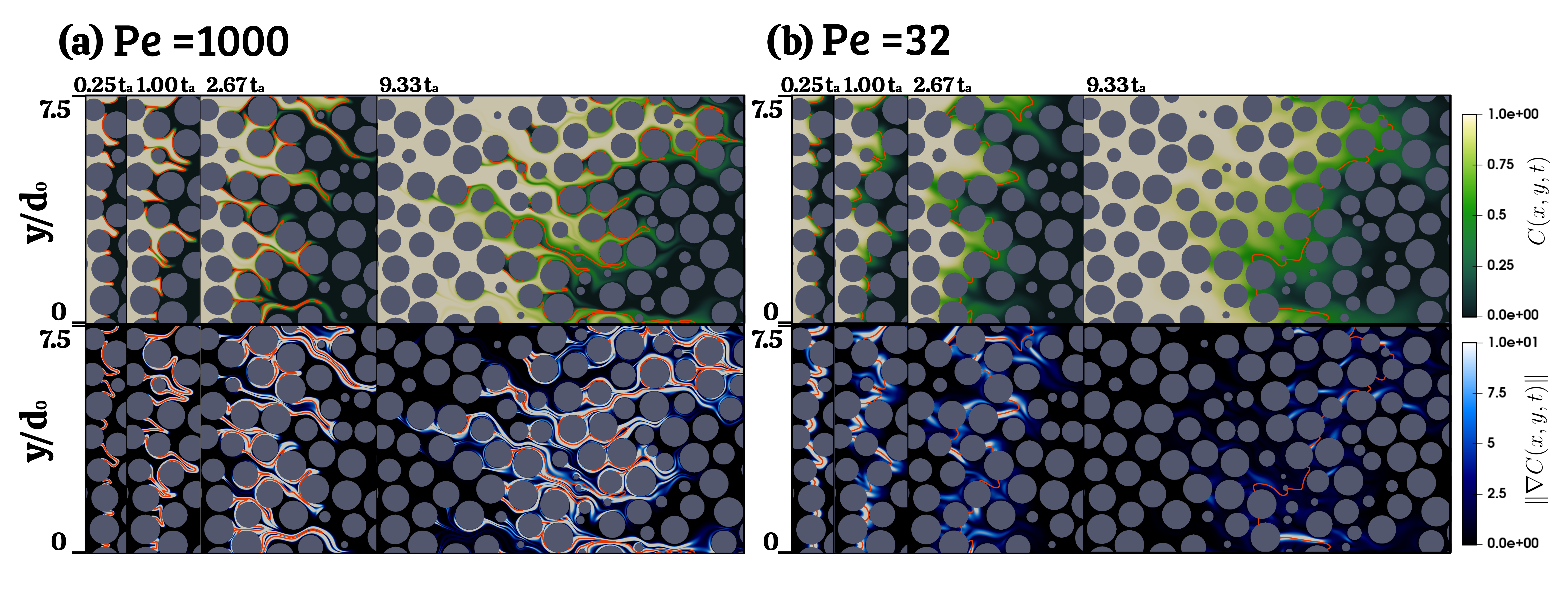}
    \caption{Portion of the domain of concentration (top) and gradient magnitude (bottom) fields at 0.25, 1, 2.67, 9.33 advection times for $Pe=32$ and $Pe=1000$. In both top and bottom, the superposed red line depicts the mixing interface ($C=0.5$).}
    \label{fig:subsample}
\end{figure}

After solving the flow, we take advantage of the system's periodicity to extend the domain length by duplicating the mesh in Figure \ref{fig:Column} (b) along the principal flow direction, thus aiming to obtain column-type scales capable of capturing asymptotic regimes of spreading and mixing processes (\cite{sole2022closer}). This approach reduces computational costs by simulating flow solely within the elementary domain. Next, the {\fontfamily{qcr}\selectfont
scalarTransportFoam} algorithm is used for simulating the transport of a conservative scalar by numerically solving the transient advection-diffusion equation,
\begin{equation}
    \label{ADE}
    \frac{\partial C} {\partial t}  + \pmb{u} \cdot \nabla C = D\nabla ^2 C \: ,
\end{equation}
where $C$ is solute concentration, and $D$ is the molecular diffusion coefficient. As the initial condition we set up a sharp front transitioning from $C=1$ to $C=0$ at  $x$-distance $d_0/2$ from the inlet (i.e., $C([x,y],t=0)=1-H(x-d_0/2)$), where $H(x)$ is the Heaviside step function. The left boundary condition is a continuous injection $C(x=0,t)=1$. The right boundary has a zero-gradient Neumann boundary condition. The simulations are stopped before the solute leaves the domain. We simulate six grain Péclet number configurations: $Pe=$ 1, 10, 32, 100, 316, and 1000, by varying the molecular diffusion coefficient, where
\begin{equation}
\label{Peclet}
Pe={\bar{u} d_0}/{D}.
\end{equation}
This range insures that we are covering transport regimes all the way from where diffusion plays a role to where advection significantly dominates. Figure (\ref{fig:subsample}) presents a portion of the domain. It shows the concentration and gradient magnitude fields for $Pe=32$ and $Pe=1000$ at different times for one realization.

\section{Temporal Evolution of the Mixing Interface and Concentration Gradients }
\begin{figure}
    \centering
    \includegraphics[width=0.7\textwidth]{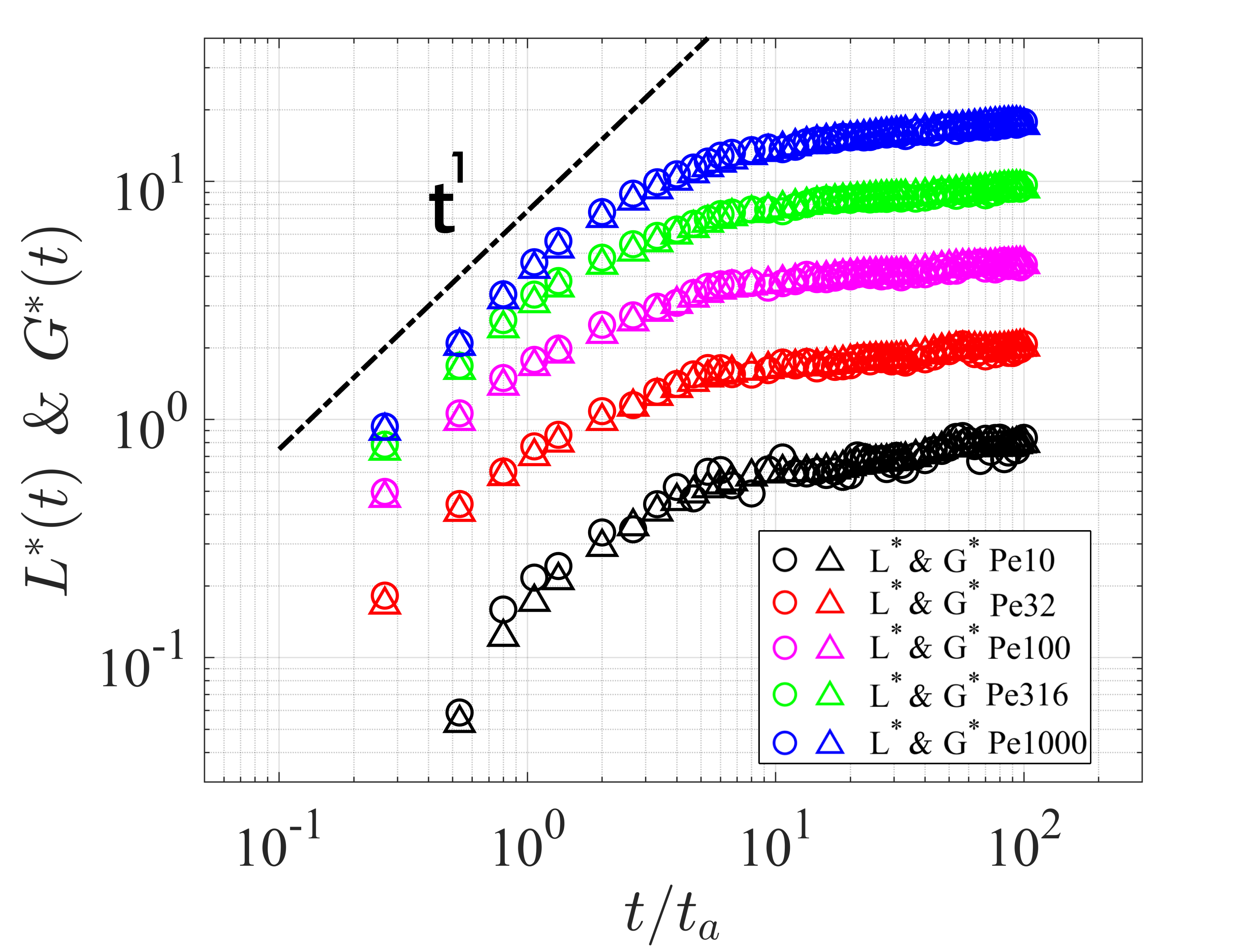}
    \caption{Ensemble of the temporal growth of the mixing interface length ($L^*$) (circles) with comparison to the growth of gradient magnitude integral ($G^*$) (triangles) across Péclet cases.}
    \label{fig:mixing_interface}
\end{figure}

In this section we present the measurements of time-dependent mixing interface length in 2D saturated random porous media as well as concentration gradient magnitudes. While these measurements are done for the transport of a single-species concentration front as described in the previous section, this setup is relevant to reactive transport as it can be re-interpreted (see Section 4.3) to represent the transport of an instantaneous bimolecular irreversible reaction.  In such a setting, the 0.5 iso-concentration line of the conservative transport simulations is analogous to the mixing interface between the reactants. The mixing interface is subjected to two main deformation mechanisms: stretching and shrinking. In two-dimensional, randomly packed porous media, the former is caused by gradients in the non-uniform velocity fields and the impact with stagnation points around grains. The latter is mainly due to diffusion, which allows solute to sample different transverse velocities, thereby limiting mixing interface growth. Lamella coalescence is another shrinking mechanism, which happens when adjacent lamellae merge to form a single, thicker layer. 
This simultaneous stretching and shrinking leads to lamellae being more elongated at higher Péclet numbers, as can be seen in Figure (\ref{fig:subsample}).  We quantify the temporal deformation of the interface, $L^*(t)$, as the relative increase in its length $L(t)$ with respect to the initial sharp interface length $L_0$.
\begin{equation}
\label{LineGrowth}
L^*(t)=\frac{L(t)-L_0}{L_0}  \: .
\end{equation}
Initially, the mixing interface undergoes rapid deformation, as shown in Figure (\ref{fig:mixing_interface}). During this ballistic regime, lengthening is dominated by differences in advection experienced by different trajectories (\cite{dentz2023mixing}). At later times, stretching and shrinking effects reach an equilibrium, resulting in a plateau in the length of the mixing interface. For Péclet number 1 and below (not shown here), the interface slightly fluctuates (due to the random variability of local pore space geometry) around its initial length, without showing any definitive growth or shrinking.

%Pore-scale concentration gradients are crucial for understanding mixing dynamics and are a fundamental measure of solute mixing. (GSM: This looks more like an introduction paragraph. We have already made clear earlier, hopefully, that our goal is to examine these two quantities and link them, and why this is important) 
At early times, a region of significant concentration gradient magnitudes is formed which aligns with, and deforms like the mixing interface, as shown in the $t_a = [0.25, 1]$ samples for Pe=1000 in Figure (\ref{fig:subsample}), $t_a=d_0/\bar{u}$ being the advection time. However, unlike the mixing interface, the region of significant gradients evolves into a two-dimensional strip-like structure, expanding over time until its connection with the mixing interface becomes less intuitive, as seen in the \(t_a = 9.33\) snapshot. Like the mixing interface length, concentration gradients tend to intensify with an increase in the Péclet number. To quantify and study the temporal evolution of concentration gradients, we calculate the change in their integral $G(t)$ over the entire
mixing area,
relative to the initial value $G_0$%=L_0$, 
\begin{equation}
\label{Gradient}
G^*(t)=\frac{G(t)-G_0}{G_0}, \,\,\,\,\, G(t)=\iint_{\Omega} 
\lVert\nabla C (x,y,t)\rVert
%\sqrt{(\nabla_x C)^2+(\nabla_y C)^2} 
\: \partial x \partial y \: .
\end{equation}
Here, $\Omega$ denotes the mixing area defined as ($\Omega = {(x,y)| \: 0 	< C(x,y) 	< 1}$). Note that due to the step initial condition the gradient field in the transverse direction is zero and a Dirac delta in the longitudinal one, meaning that $G_0=L_0$. At early times, $G^*(t)$ increases because the concentration gradients at the pore-scale decay slower than the rate at which they are created as the plume spreads. Later on, both the decay of concentration gradients and the plume spreading rate converge to an equilibrium, attaining a constant value for $G^*(t)$. Our simulation results reveal that $G^*(t)$ aligns precisely with the growth of the mixing interface $L^*(t)$, as shown in Figure (\ref{fig:mixing_interface}), not only initially and at early times but for the full duration. The interface length at any given moment is, in fact, the same as integral of the pore-scale gradient magnitudes over the entire mixing area despite the seeming qualitative divergence at later times discussed in relation to Figure 2. This result further highlights the relevance of measuring and predicting the mixing interface length, as it can be directly linked to concentration gradients and therefore mixing and reaction.
%In fact, since as noted earlier $G_0=L_0$, these results point out to a direct equality between the mixing interface length \(L(t)\) and the integral of gradient magnitudes \(G(t)\) (note that this assumes a unitary solute concentration in the invading solution).

%\(G_0\) can be simplified as follows: (a) at time 0, $\lvert \nabla_y C \rvert=0$. Gradient magnitudes, which are perpendicular to the isolines, exist only in the main flow direction. (b) Gradient magnitudes in the principal flow direction are equal to $\lvert \nabla_x C \rvert = \mathds{1} \cdot \delta(x-d_0/2)$ over \(y \in [0, H]\).
%Here, \(\mathds{1}\) is the indicator function and is equal to 1 for water filled pores and 0 otherwise. $\delta(x)$ is the Dirac delta function (i.e. the derivative of the Heaviside step initial condition). $H$ is the domain width. Therefore, $G_0=nH=L_0$.  
%This point out to a direct equality between the mixing interface length \(L(t)\) and the integral of gradient magnitudes \(G(t)\) (note that this assumes a unitary solute concentration in the invading solution).

%This presents a potential advantage as experimentally it may be easier to measure one over the other. Also, we find this insight valuable because understanding how the mixing interface, a local quantity, deforms provides us with insights into the behavior of the mixing area/volume and vice versa. For instance in a two dimensional plug flow with a continuous injection, all solutes within the flow have identical velocity and direction of motion and the sharp initial interface remains undeformed.  From $L(t) = G(t)$, one can infer that the integral of the local gradient magnitudes will not change either. Thus, local gradients would decay at the same rate as plume spreading. This is consistent with such a system as the gradients are only in the flow direction and the maximum gradient decays as $\nabla C(x,t) \sim 1/\sqrt{4\pi D t}$.  Meanwhile, the plume width spreads as $\sqrt{2Dt}$.
%Consider now another endmember case: a purely advective Poiseuille flow with no diffusion. Here the mixing interface length would grow indefinitely, necessitating $G(t)$ to grow indefinitely also. The concentration difference across the front, $\Delta C$, has to remain constant, requiring the compression of the solute front width, $s(t) \sim t^{-1}$ to precisely match the growth of $L(t) \sim t$. This occurs due to incompressibility of the fluid and agrees with \cite{ranz1979applications}.
\section{Modeling the Link of Concentration Gradients and Mixing Interface Length}
Motivated by, and building upon the findings of the previous section, we aim to write a direct mathematical expression for predicting mixing interface length, which can then be used to estimate mixing-limited reactive transport. The model exploits a local link between the orientation (slope) of the mixing interface and transverse concentration gradients. We ultimately show that the identity found numerically in the previous section (Figure 3) can be explained by this analytical model. We start with a classical simple pressure-driven flow in a two-dimensional channel (Poiseuille flow). Although this setting lacks critical defining features of flow in porous media%such as tortuosity
, it can often serve as an idealized representation of flow through individual pores %in more complex porous media
(\cite{de2017prediction,al2024effects}). We derive an analytical solution for the asymptotic time behavior and then tackle the transient regime. Finally, we combine all of this to propose an analytical solution to estimate the upscaled reaction kinetics in porous media, accounting for incomplete mixing.
\subsection{\fontsize{12}{14}\selectfont  Asymptotic Regime in a 2D Poiseuille Flow}
\begin{figure}
    \centering
    \includegraphics[width=\textwidth]{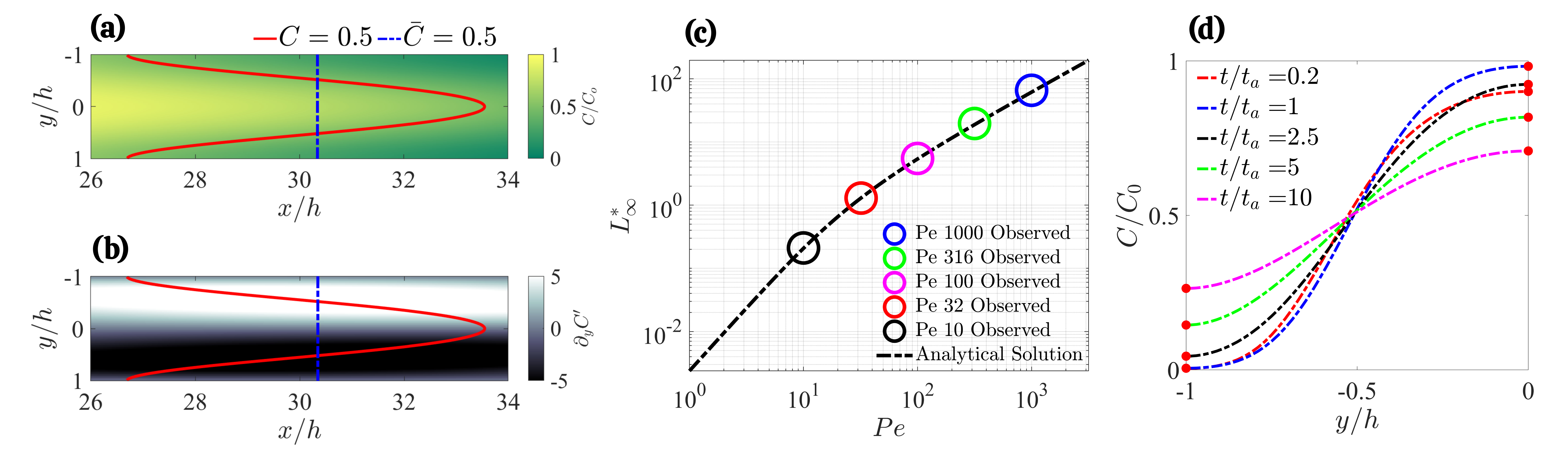}
    \caption{(a) Scalar field for 2D Poiseuille flow at $Pe=100$ and 10 advection times, showcasing the 0.5 mixing interface isoline (in red) and 0.5 mean concentration location (in blue); (b) same as (a) but showing the transverse gradient of concentration fluctuations; (c) quantification of the plateau mixing interface growth across various Péclet numbers, comparing observations to equation (\ref{LenPlat}); (d) Concentration profile along $\mu_1(t)$ in 2D Poiseuille flow simulation for $Pe=100$ at different times.}
    \label{fig:Analytical}
\end{figure}

The velocity \(u(y)\) of a Newtonian-fluid's pressure-driven laminar flow between two parallel plates separated by a distance \(2h\), subjected to no-slip boundary conditions, is given by
\begin{equation}
\label{velocity}
u(y)=\frac{3}{2}\bar{u}(1-\frac{y^2}{h^2}) \qquad -h\le y\le h \: .
\end{equation}
A snapshot of the concentration field and the mixing interface after 10 advection times for $Pe=100$ is presented in Figure \ref{fig:Analytical}(a). Starting from the standard two-dimensional advection-diffusion equation for a conservative scalar \(C(x, y, t)\), we use Reynolds decomposition to separate \(C(x, y, t)\) and \(u(y)\) into their respective means ($\bar{C}(x,t)$, $\bar{u}$) and fluctuations ($C'(x,y,t)$, $u'(y)$). At asymptotic times (i.e. after a characteristic diffusive time scale $t_D=\frac{(2h)^2}{D_m}$), we can invoke Taylor dispersion theory (\cite{taylor1953dispersion}) to derive an analytical expression of the mixing interface slope $m(y)$ at each point along the $y$-axis. The derivation is outlined in Appendix A.  Using this approach, we can explicitly link the local interface slope to concentration gradients as: 
\begin{equation}
\label{slope}
m(y)=\frac{\partial x}{\partial y}=-\frac{\nabla_y C'(y)}{\nabla_x \bar{C}}\big|_{x=\mu_1(t)}=\frac{\bar{u}}{2D}(y-\frac{y^3}{h^2}) ,
\end{equation}
where $\nabla_y C'(y)$ is the transverse gradient of concentration fluctuations and $\nabla_x \bar{C}$ is the longitudinal gradient of mean concentrations, both calculated at the location of the longitudinal first spatial moment ($\mu_1(t)=\bar{u}t+x_0$) of the mixing interface (i.e. $\bar{C}=0.5$) (see Figure \ref{fig:Analytical}(b)). Equation \ref{slope} directly proves that the transverse variation in the concentration field ($\nabla_y C'(y)$) under Poiseuille flow never converges to a uniform concentration. Although it will decrease over time it persists in a balance with the mean concentration gradient forever and can be important for reactive transport.

\begin{comment}
In the context of this specific transport setting and initial condition, $\mu_1(t)$ can be defined by interpreting the one-dimensional mean concentration $\bar{C}(x,t)$ as the complementary cumulative density function in $x$ for the transport of a pulse initial condition ($\bar{C}_{\delta}$).
Assuming that the mixing area/volume is within the column, the first spatial moment of \(\bar{C}_{\delta}(x,t)\) can be calculated using \(\bar{C}(x,t)\) as 

\begin{equation}
\label{moments}
\mu_1(t)=x_{0}+\frac{1}{C_0}\int_{x_{0}}^{x_{outlet}} \bar{C}(x,t) \partial x %\quad \quad m_{(i)}(t)=[x_{in}-\mu_1(t)]^i + \frac{i}{C_0}\int_{x_{in}}^{x_{out}} (x-\mu_1)^{i-1} \bar{C}(x,t)\partial x \: ,
\end{equation}
where $x_{0}$ is the location of the initial sharp interface at $t=0$.
\end{comment}

With the formulation for the slope given in equation (\ref{slope}), the plateau mixing interface length growth ($L^*_{\infty}$) relative to the initial length can be calculated using the arc length formula as:
\begin{equation}
\label{LenPlat}
L^*_{\infty}(Pe)=\frac{\int_{-h}^{h}\sqrt{1+\left( \frac{\partial x}{\partial y}\right)^2}\partial y -L_0}{L_0} =    \int_0^1\sqrt{\frac{Pe^2}{16}(y^*-y^{*^3})^2+1} \, \, \partial y^* \, -1 \: ,
\end{equation}
where $y^*=y/h$. In Figure \ref{fig:Analytical}(c), we validate our analytical solution by numerically integrating equation \ref{LenPlat} and comparing it to the measured $L^*_{\infty}$ for a set of two-dimensional Poiseuille flow simulations. We observe an exact match between the observed $L^*_{\infty}$ and the theoretical one. 
Hence, in diffusion dominated systems ($Pe\lesssim 1)$, the interface length experiences no significant growth ($L^*_{\infty}\approx 0$). On the other hand, in advection dominated systems, characterized by high Péclet numbers, $L^*_{\infty}$ scales as $Pe/16$.
%We note that equation \ref{LenPlat} was derived for Poiseuille flow, where streamlines maintain a constant velocity.
The key thing to highlight here is that mixing interface deformation directly stems from (i) the gradient of transverse concentration fluctuations (see Figure \ref{fig:Analytical}(b)) and (ii) the gradient of the mean concentration in the longitudinal direction. We will continue to assume this to model the transient behaviour.

\subsection{\fontsize{12}{14}\selectfont  Transient Regime}
At time $t = 0$, the initial sharp mixing interface has a zero slope, i.e., $m(y, t = 0)=0$. Subsequently, it undergoes simultaneous steepening and flattening. %due to velocity shear, molecular diffusion, and coalescence (in porous media).
At late times, and under Poiseuille flow setting, $m(y, t \rightarrow \infty)$ becomes constant and is described by equation (\refeq{slope}). However, for flow in porous media, $m(y, t \rightarrow \infty)$ fluctuates, but its mean along the mixing interface, $\bar{m}(t)$, eventually converges to a constant value as its length reaches a plateau. To mathematically describe this, we aim to model the gradients across a single lamella with a characteristic width $2h_c$ by quantifying the temporal evolution of (i) $\nabla_y C'(y,t)$  and (ii) $\nabla_x \bar{C}(t)$ at \(x=\mu_1(t)\). 

To begin, we first assume that the mixing isoline has a continuous and differentiable functional form $x_L=f_L(y,t)$. By definition, the length of this isoline can be expressed using its first spatial derivative ($\partial_y f_L(y,t)$) and the gradients along it as
\begin{equation}
\label{Length}
L(t)=\int_0^{H} \sqrt{1+ \left[ \frac{\partial f_L(y,t) }{\partial y} \right]^2 } \, \partial y = \int_0^{H} \sqrt{1+ \left[- \frac{\nabla_y C(y,t)}{\nabla_x C(y,t)} \right]_{iso} ^2 } \, \partial y \: .
\end{equation}
To simplify (\ref{Length}), we approximate the concentration gradients along the 0.5 mixing isoline using those at the first spatial moment ($x=\mu_1(t)$), as supported by the proof in Appendix A. Although the proof is derived for asymptotic times, we assume that it is a valid approximation during the transient regime also, as we will demonstrate later.   With this in mind, we take $H$ as the channel width for Poiseuille flow, and the water-filled voids for porous media (i.e. $nH$). Within a single lamella,  $\nabla_x C(y,t)$ along the 0.5 isoline can be approximated by its mean value $\nabla_x \bar{C}(t)$. Hence, the interface length calculation can be simplified as  
\begin{equation}
\label{LengthAprox}
L(t) \approx \int_0^{H} \sqrt{1+ \left[- \frac{\nabla_y C'(y,t)}{\nabla_x \bar{C}(t)} \right] _{\mu_1(t)}^2} \, \partial y \approx H \sqrt{1+ \left[- \frac{\nabla_y C'(\zeta,t)}{\nabla_x \bar{C}(t)} \right] _{\mu_1(t)}^2} =  H \sqrt{1+ \left[ - \frac{\overline{\nabla_y C'}(t)}{\nabla_x \bar{C}(t)} \right]_{\mu_1(t)}^2} \: .
\end{equation}

Here we have used the mean value theorem for integrals to quantify the mixing interface length in terms of the mean slope $\bar{m}(t)=-{\overline{\nabla_y C'}(t)/{\nabla_x \bar{C}}(t)}$. $\zeta$ is an unspecified point along $x=\mu_1(t)$ at which the value of $\nabla_y C'(\zeta,t)$ is equal to the mean value of the function $\nabla_y C'(y,t)$ over $[0 \: h_C]$. Thus, $\nabla_y C'(\zeta,t)$ can be replaced by $\overline{\nabla_y C'}(t)$ which is defined as $\overline{\nabla_y C'}(t) = (1/h_C)\int_0^{h_C} \nabla_y C'(y,t) \: \partial y$.

A sample of the temporal evolution of the concentration profile along $x= \mu_1(t)$ for the Pe=100 Poiseuille flow simulation at different times is presented in Figure \ref{fig:Analytical}(d). From equation (\ref{LengthAprox}), one only needs to quantify the temporal change of concentrations fluctuations at the maximum mean longitudinal velocity $C'(t,y_t=0)_{\mu_1(t)}$ and the minimum one $C'(t,y_t=h_C)_{\mu_1(t)}$ over a half lamella width ($h_c$) to estimate $\overline{\nabla_y C'}(t)$. In other words, we need to find the difference between the red dots shown in Figure \ref{fig:Analytical}(d) along each line. Hence, the mean transverse gradient can be quantified as:
\begin{equation}
\label{TransvGrad}
\overline{\nabla_y C'}(t)_{\mu_1(t)}=\left[ \frac{C'(t,y_t=0)-C'(t,y_t=h_C)}{h_C} \right]_{\mu_1(t)} \: .
\end{equation}
Here, we define the lamella width as the characteristic distance between the peak and trough concentrations along $x=\mu_1(t)$. A useful analogy here is to think of the concentration profile along $x=\mu_1(t)$ as a wave, with the lamella width representing the wavelength, where $C'(t,y_t=0)_{\mu_1(t)}$ and $C'(t,y_t=h_c)_{\mu_1(t)}$ are the peak and trough values.  In a Poiseuille flow, the lamella width ($2h_c$) is equal to the channel width ($2h$) and remains constant over time. It is independent of the Peclet number due to lack of folding, and coalescence.
 
\begin{figure}
    \centering
    \includegraphics[width=\textwidth]{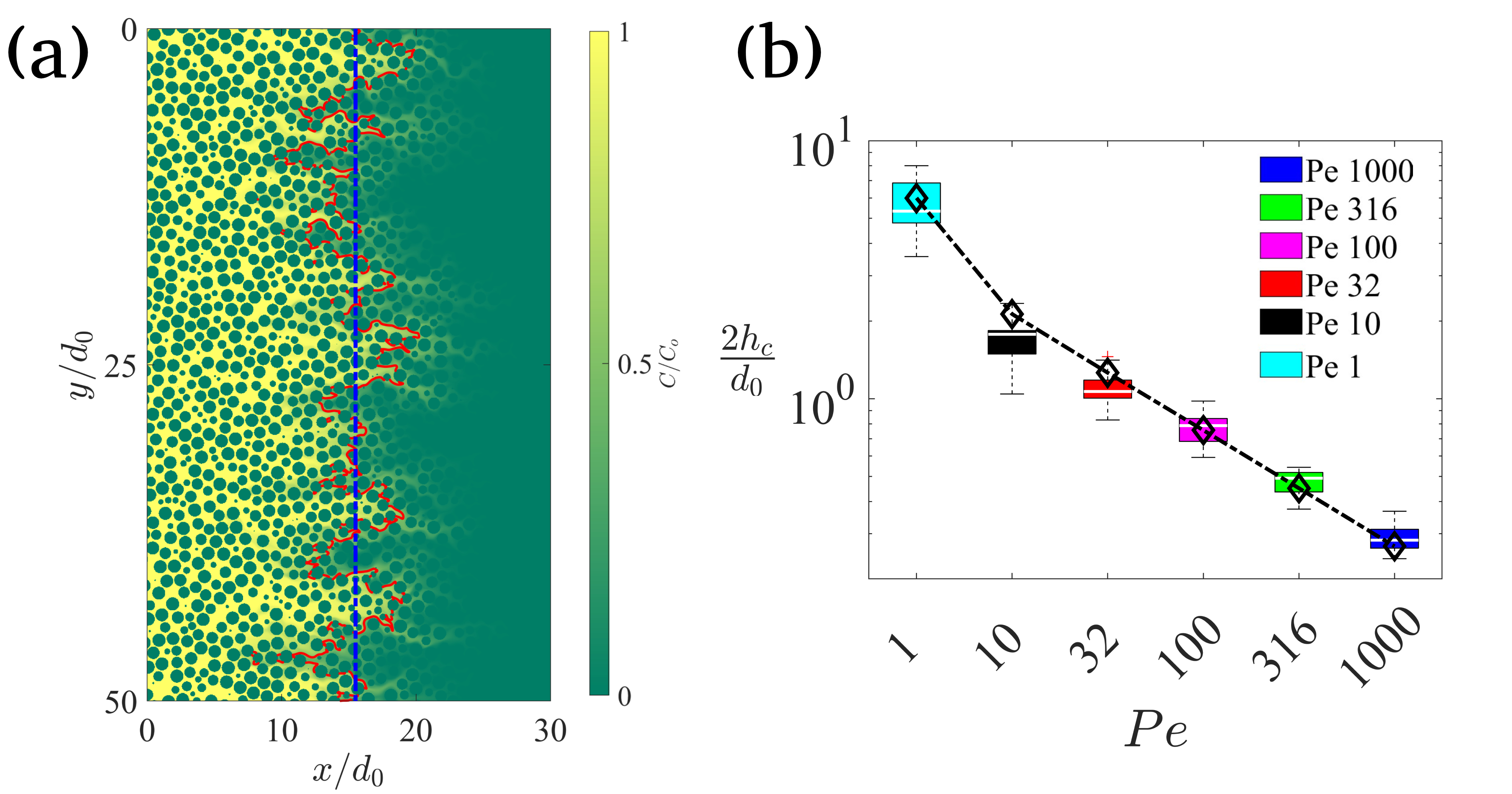}
    \caption{(a) Scalar field for 2D porous media flow with $Pe = 100$ at 20 advection times, highlighting the 0.5 mixing interface isoline in red and the 0.5 mean concentration location in blue. (b) Box plot of the normalized lamella width, \(2h_c\), at late times, quantified by the number of intersections, \(N\), between the continuous 0.5 isoline and the 0.5 mean concentration location (i.e., \(2h_c \approx nH /N \)). Results are shown for different Péclet numbers over all five realizations and compared to the width described in Equation \eqref{lengthscale}, represented by a black dashed line.}
    \label{fig:Intercepts}
\end{figure}

To quantify this lamella width in randomly packed porous media, we begin by assuming that the mixing interface can be represented as an ensemble of idealized lamellae, each having a $2h_c$ width. Thus, it is sufficient to measure the elongation of a single representative lamella to estimate the deformation of the entire mixing interface. At early times, this lamella’s width is determined by the characteristic pore size and is Peclet independent. However, at later times, due to processes like folding, stirring, and coalescence, the width asymptotes to a constant value that is Peclet dependent. To estimate this width ($2h_c$), we redefine the Peclet number in equation (\ref{LenPlat}) as $Pe_c = \frac{u(2h_c)}{D_m}$. By fitting $L^*_{\infty}(Pe_c)$ to the observed plateau length growth, as shown in Figure (\ref{fig:mixing_interface}), we can estimate the characteristic lamella width as
\begin{equation} 
\label{lengthscale} 
2h_c \approx \frac{4}{3}Pe^{-0.45} . 
\end{equation}
The higher the Peclet, the smaller the lamella width due to stronger folding and weaker coalescence. We cross-checked this width by counting the characteristic lamella width directly from the simulations. In Figure \ref{fig:Intercepts} (a), we show an example of the concentration field for Pe = 100 at 20 advection times,
highlighting the 0.5 mixing interface isoline in red and the 0.5 mean concentration location in blue. From this, $2hc$ can be quantified by dividing the water-filled pore width by the number of intersections ($N$) between the continuous 0.5 isoline and the 0.5 mean concentration location (i.e., $2h_c \approx nH/N$ ). We did this for the simulated Peclet range at late times ($t/t_a>20$) across all five realizations. In Figure \ref{fig:Intercepts} (b), we show a boxplot for the acquired length scales, which are consistent with the value defined in equation (\ref{lengthscale}).

We assume that the width defined in equation (\ref{lengthscale}) is valid for transient times too. This has negligible effects after one advection time, as we will demonstrate later.  It is interesting to note that the width of the concentration fluctuations ($2h_C$) scales with the Peclet number in a manner similar to the Batchelor scale for exponential stretching ($S_B \propto Pe^{-1/2}$), rather than the scaling observed for power-law stretching ($S_B \propto Pe^{-1/3}$) (\cite{souzy2018mixing, villermaux2019mixing}). This suggests that the stretching rate in a two-dimensional randomly packed porous medium is stronger than that of a simple shear flow. However, further investigation of this will be the focus of future work. For the current study, we will continue using the lamella width as defined in equation (\ref{lengthscale}).

To quantify $\bar{m}(t)$, we propose two analytical approaches and give the reader the liberty to choose between them 

(1) by separately calculating the transverse concentration gradient $\overline{\nabla_y C'}(t)$ and the longitudinal concentration gradient $\nabla_x \bar{C}(t)$, where their ratio yields $\bar{m}(t)$. The concentration profile at a given location $y_t$ is modeled using a complementary error function, assuming an apparent velocity $u_a(t)$ and longitudinal dispersion $D_l(t)$. The apparent velocity is computed by convolving the position distribution $P(y,t)$, obtained from the Green's function, with the longitudinal velocity $u(y)$. The transverse gradient is then calculated from the difference in concentration fluctuations at two transverse positions ($y_t=0$ and $y_t=h_c$). For the longitudinal gradient, it can be estimated using the maximum gradient of the classical advection-diffusion equation with a continuous injection. For full details, we direct the reader to Appendix B.

(2) using the closure solution for $C'(x,y,t)$ proposed by (\cite{bolster2011mixing}). They applied volume averaging to estimate the concentration field of a conservative solute in a stratified flow from early to late times by relaxing several of the assumptions required for Taylor dispersion, only valid at late times. Their proposed closure is
\begin{equation}
\label{DiogoCfluc}
C'(x,y,t)=b_0(x,y,t)+b_1(y,t)\nabla_x \bar{C}(x,t) \: ,
\end{equation}
where $b_0(x,y,t)$ and $b_1(y,t)$ represent the influence of the initial condition and convective source, respectively.
For the considered Heaviside initial condition, the former is zero (see \cite{bolster2011mixing} for full details), while the latter is given by
\begin{equation}
\label{b1}
b_1(y,t)=- \int_0^t \int_0^{h_C} u'(\eta) Gr(y,\eta,t-\tau)\partial \eta \partial \tau \: ,
\end{equation}
where $Gr(y,\eta,t-\tau)$ is the Green's function for transverse diffusion across the domain, given by (\cite{polyanin2001handbook})
\begin{equation}
\label{green}
Gr(y,\eta,t-\tau)=\frac{1}{h_C} + \frac{2}{h_C} \sum_{\alpha=1}^{\infty} cos(\frac{\alpha \pi y}{h_C}) cos(\frac{\alpha \pi \eta}{h_C}) e^{-\frac{D\alpha^2 \pi^2 (t-\tau)}{h_C^2}} \: \quad \quad  0\leq y \leq h_C \: .
\end{equation}
The velocity fluctuation $u'(\eta)$ can be found by subtracting the mean longitudinal velocity ($\bar{u}$) from equation (\ref{velocity}).
Thus, executing the integrals in \eqref{b1} we can quantify the concentration fluctuations at  $C'(t,y_t=0)_{\mu_1(t)}$ and $C'(t,y_t=h_C)_{\mu_1(t)}$ and we get the following analytical solution for the mixing isoline mean slope:
\begin{equation}
\label{b1_2}
\bar{m}(t)=-\left[ \frac{\overline{\nabla_y C'}(t)}{\nabla_x \bar{C}(t)} \right]_{\mu_1(t)}=\frac{6\bar{u}h_C}{\pi^4D}\sum_{\alpha=1}^{\infty} \frac{(1-(-1)^{\alpha})(1-e^{-4\alpha^2\pi^2 \frac{t}{t_D}})}{\alpha^4} \: ,
\end{equation}
When $t/t_D>>1$, diffusion has had sufficient time to significantly spread the scalar over the characteristic distance $2h_C$ and the mean slope of the interface becomes constant. At late times $(t\rightarrow \infty)$, equation (\ref{b1_2}) matches the mean of the slope described earlier in equation (\ref{slope}).

\begin{figure}
    \centering
    \includegraphics[width=\textwidth]{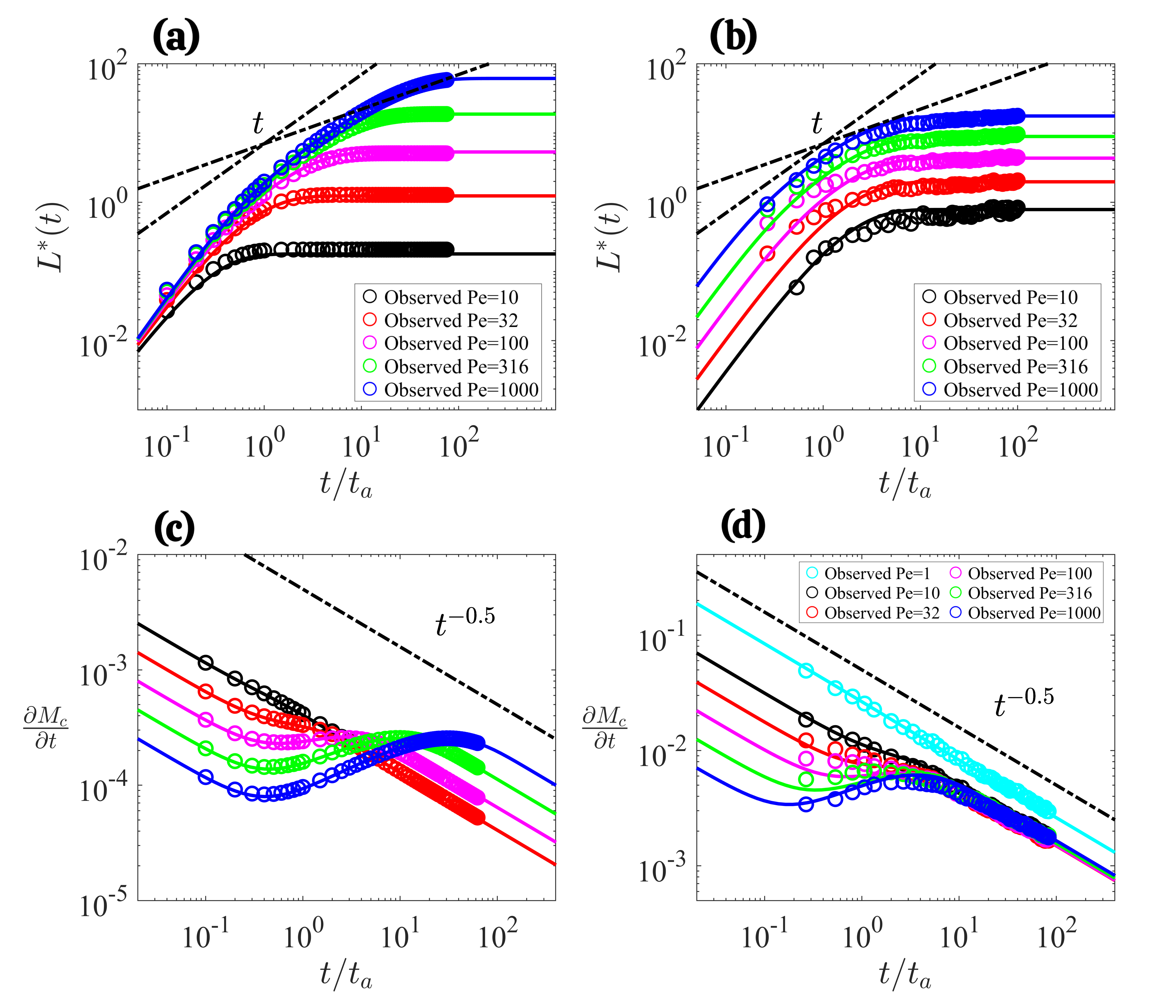}
    \caption{Comparing the growth of the mixing interface length between simulations (circles) and the analytical solution (solid lines) across different Péclet numbers in (A) Poiseuille flow and (B) porous media. This also presents a comparison for the rate of \(C\) production in (C) Poiseuille flow and (D) porous media settings.}
    \label{fig:AnalyticalTransient}
\end{figure}

In Figure \ref{fig:AnalyticalTransient} (a), we show the close alignment between the mixing interface elongation quantified using our analytical solution to that from the set of Poiseuille flow simulations for different Peclet numbers. %A negligible underestimation is observed simply because the length is calculated using the mean slope $\bar{m}(t)$ rather than $m(y, t)$. 
Next, we apply this approach to estimate $L^*(t)$ in porous media, only this time using the characteristic length scale $2h_C$ as prescribed in equation (\ref{lengthscale}). 
Figure \ref{fig:AnalyticalTransient} (b), show the comparison for the porous media, and it matches nicely with the observation for $\sim \frac{t}{t_a}\geq 1$. This early time mismatch is unsurprising prior to one advection time, mainly because the characteristic length scale of concentrations fluctuations at such early times is not well represented by the one prescribed in equation (\ref{lengthscale}) (or the characteristic lamella width has not had time to evolve yet). This is also why lines do not converge as time approaches zero ($t \rightarrow 0$), in a manner similar to Figure \ref{fig:AnalyticalTransient} (a). However, we find it interesting that after as little as one advection time (the characteristic time to traverse one grain), the difference between a complex porous media flow and a simple Poiseuille flow can all be encapsulated within the characteristic width scale ($2h_C$).

Lastly, we make use of the theory developed in this section to link back to the data presented in Section 3. In Appendix C, we show that the model predicts the equality between the mixing interface length, $L(t)$, and the integral of pore-scale concentration gradients, $G(t)$,
%By expressing $G(t)$ in terms of the concentration gradient components and the interface slope $m(x, y, t)$, we derive an explicit formula for $G(t)$, which matches the approximation for $L(t)$. This confirms that the mixing interface length is indeed equal to the integral of concentration gradient magnitudes across the domain, [redundant]
supporting our observations in Figure \ref{fig:mixing_interface}). 

\subsection{\fontsize{12}{14}\selectfont  Temporal Evolution of Mixing-Limited Reactive Transport}
The conservative transport simulations can be re-interpreted to study an instantaneous, bimolecular, irreversible
reaction ($C_A+C_B \rightarrow C_C$).
The subscripts, A, B, and C refer to the invading solution, the displaced one, and the reaction product.  To do so, one can define two reaction independent, conservative species $u_A$ and $u_B$, as
\begin{equation}
\label{stoch}
u_A=C_A+C_C  \,\,\,\,\,\,,\,\,\,\,\,\, u_B=C_B+C_C \: .
\end{equation}
Assuming an instantaneous, irreversible reaction, $C_A$ and $C_B$ cannot coexist at the same point in space and time; thus the concentration of $C_C$ can be defined as:
\begin{equation}
\label{Cc}
C_C(x,y,t)=min(u_A(x,y,t),u_B(x,y,t)) \: .
\end{equation}
This assumes that all chemical species have the same molecular diffusion coefficient.
We now aim to propose a solution for the production rate of $C_C$ in both Poiseuille flow and porous media flow.  In mixing limited reactive transport, the production of $C_C$ is limited by the flux of either $C_A$ or $C_B$ across the interface between the two solutions. For a spatially homogeneous system, the product concentration $C_C$ is symmetric around the mixing interface (\cite{gramling2002reactive}). Hence, the rate of $C_C$ production can be represented by integrating Fick's law along the mixing interface as:
\begin{equation}
\label{rateProduct}
\frac{dM_c(t)}{dt}=2 \int_{L(t)} D \left| \nabla C_A(\xi,t)  \right| dl \approx 2DL(t)\overline{\left| \nabla C_A(t) \right|} \: .
\end{equation}

Here, $\xi$ represents the coordinate normal to the mixing interface. The average gradient magnitude along the 0.5 isoline ($\overline{\left| \nabla C_A(t) \right|}$) can be estimated in terms of the averages of the pore-scale longitudinal ($\overline{\left| \nabla_x C_A(t) \right|}_{iso}$) and transverse ($\overline{\left| \nabla_y C_A(t) \right|}_{iso}$) concentration gradients along the mixing interface. %The longitudinal gradients and their average along the 0.5 isoline (i.e. ${\left| \nabla_x C_A(y,t) \right|}_{iso}$, and $\overline{\left| \nabla_x C_A(t) \right|}_{iso}$) decay as $\sim 1/{\sqrt{4\pi D_e(t) t}}$, in which $D_e(t)$ is the effective dispersion coefficient. 
The average longitudinal gradient along the 0.5 isoline decay as $1/{\sqrt{4\pi D_e(t) t}}$, in which $D_e(t)$ is the effective dispersion coefficient. The effective dispersion coefficient $D_e(t)$ measures the mean width of a point injection within the channel's cross section, essentially representing the Green's function of the transport problem.  In appendix D, we show that the effective dispersion coefficient can be directly quantified using the mixing interface mean slope $\bar{m}(t)$ as:
\begin{equation}
\label{effecDisp}
D_e(t)=D(1+1.25\bar{m}^2(t)) .
\end{equation}
This means that the effective dispersion could be directly approximated from the length of the mixing interface as $D_e(t)/D \approx 1.25(L(t)/L_0)^2$.
Lastly, the average transverse concentration gradients can be quantified using the longitudinal one as $\overline{\left| \nabla_y C_A(t) \right|}_{iso} =\bar{m}(t)\overline{\left| \nabla_x C_A(t) \right|}_{iso} $ . Hence, we can re-write equation (\refeq{rateProduct}) as
\begin{equation}
\label{rateProduct2}
\frac{dM_c(t)}{dt}= L_0\sqrt{\frac{D}{\pi t}}\frac{1+\bar{m}^2(t)}{\sqrt{1+1.25\bar{m}^2(t)}} .
\end{equation}
Here, $L_0$ represents the initial length of the mixing interface. For Poiseuille flow, $L_0$ is equal to $H$, while in porous media, it is $nH$. In Figure \ref{fig:AnalyticalTransient} (c), we present a comparison between the observed rate of mass production from the simulations and the analytical solution outlined in this paper, showing an exact alignment between the two. For $t/t_D>>1$, both $D_e(t)$ and $\overline{m}(t)$ converge to constant values and equation (\ref{rateProduct2}) scales as $\propto t^{-1/2}$.
We use equation \eqref{rateProduct2} to specify the following two limiting regimes approximations: 
\begin{enumerate}
    \item For diffusion-dominated systems, $\frac{dM_c(t)}{dt} \approx L_0 \sqrt{D/(\pi t)}$.

    \item For advection-dominated systems, $\frac{dM_c(t)}{dt} \approx L_0 \sqrt{D/(4t)} \: \bar{m}(t) $.
    
\end{enumerate}
Our diffusion-dominated, limiting regime approximation is equivalent to the well-mixed solution proposed by \cite{gramling2002reactive}. 
\begin{equation}
\label{McGramling}
\left.\frac{dM_c(t)}{dt}\right|_{w m}=L_0\sqrt{D^*/(\pi t)} \: ,
\end{equation}
where $D^*$ is the sum of the diffusion and hydrodynamic dispersion coefficients which for $\sim Pe \leq 1$ is approximately equal to the molecular diffusion  (\cite{delgado2007longitudinal}).
However, for $Pe > 1$, it is no longer a valid assumption and our model demonstrates how to account for incomplete mixing.  In Figure \ref{fig:AnalyticalTransient} (d), we demonstrate how our model closely captures the observed data under porous media flow.

If one normalizes the reaction production rate in equation (\ref{rateProduct2}) by the plume spreading length, we get:
\begin{equation}
\label{normalized}
\frac{C_C \: Production \: \: rate}{Spreading \: \: length}: \frac{1}{\sqrt{2D_e(t)t}}\frac{dM_c(t)}{dt} \approx L_0\frac{1}{\sqrt{2\pi}t} \: .
\end{equation}
This implies that the temporal production rate of $C_C$ per unit mixing area is independent of the Péclet number and is approximately \( (\sqrt{2\pi}t)^{-1} \).  It is also worth highlighting that the late-time $C_C$ production rate in porous media shows substantially less variability than under Poiseuille flow over the simulated Péclet range (see Figure \ref{fig:AnalyticalTransient}). This weak Péclet scaling could be linked to the inverse dependence of the characteristic lamella width scale on Péclet number. For $Pe>1$, and using equations (\ref{lengthscale}), (\ref{b1_2}), and (\ref{rateProduct2}), we get this weak dependence for late times as $ \propto Pe^{0.05}\:$, which closely matches the observed $\sim Pe^{0.035}$ scaling.
\section{Summary and Conclusion}
In this paper, we conducted a set of high-resolution pore-scale flow and transport simulations to study mixing-limited reactive transport under laminar saturated flow conditions. Using OpenFOAM, we solved the Navier-Stokes and advection-diffusion equations for a periodic porous medium generated by a packing algorithm that randomly arranges spherical grains. Our simulations cover a wide range of Peclet numbers, from diffusion-dominated to advection-dominated transport regimes. We observed that under Poiseuille flow, and in fully saturated porous media, the mixing interface length and the integral of pore-scale concentration gradients undergo an initial growth until an equilibrium between stretching and shrinking is reached.% At this point, neither of them elongate nor shrink.
We observed that the length of the mixing interface at any moment is equal to the integral of the pore-scale concentration gradient magnitudes over the entire mixing area/volume. This presents a potential advantage, as experimentally it may be easier to measure one over the other. Additionally, we find this insight valuable because understanding the deformation of the mixing interface, a local quantity, provides insight into the behavior of the overall mixing area or volume, and vice versa.

For example, in a two-dimensional plug flow with continuous injection, where all solutes have identical velocities, the mixing interface length $L(t)$ remains constant. Consequently, one could also infer that the integral of the local concentration gradient magnitudes $G(t)$ is constant as well. Thus, local gradients would decay at the same rate as plume spreading. This can be confirmed by noting that in such a system, where gradients exist only in the flow direction, the maximum gradient decays as $\propto 1/\sqrt{4\pi D t}$. Meanwhile, the plume width spreads as $\propto \sqrt{2Dt}$. Now, consider another endmember case: a purely advective Poiseuille flow with no diffusion. Here the mixing interface length would grow indefinitely, necessitating $G(t)$ to grow indefinitely as well. Since the concentration difference across the front must remain constant, the solute front width would need to compress at a rate of $\propto t^{-1}$, precisely matching the growth of $L(t) \propto t$. This behavior is due to the incompressibility of the fluid and is consistent with the findings of \cite{ranz1979applications}.

The observation that the mixing interface length reaches a plateau without subsequent decay somewhat challenges the classical interpretations associated with Taylor dispersion theory, which state that transverse concentration gradients vanish at late times. To address this potential discrepancy, we developed and validated a theoretical model that quantifies the deformation of the mixing interface. This is done by estimating the decay of both longitudinal and transverse concentration gradients along the 0.5 mean concentration contour. The model is initially derived for an idealized Poiseuille flow and later extended to apply to flow in porous media by introducing a characteristic width scale for pore-scale concentration fluctuations.

Our results show that this solution closely aligns with observations in both flow regimes. We show that transverse concentration gradients persist indefinitely and are crucial for accurately modeling reactive transport. Specifically, for $Pe > 1$, incomplete mixing, once established, persists. This provides a new perspective on mixing-limited reactive transport in both pipe flow and porous media under laminar flow conditions. Finally, based on the theory presented, we propose a model for estimating upscaled reaction kinetics. For $Pe \leq 1$, the well-mixed assumption provides a reliable estimate of mixing-limited reactive transport. However, for $Pe > 1$, this assumption no longer holds, and our model shows how to account for incomplete mixing.

\printbibliography

\section*{Appendix A: Mixing Interface Plateau Length in Poiseuille Flow}
\renewcommand{\theequation}{A.\arabic{equation}} 
\setcounter{equation}{0} % Reset equation counter

In this appendix we derive an analytical solution for the mixing interface plateau length in Poiseuille flow. We derive it for the case of flow between parallel plates, but the same solution could be retrieved if circular pipe flow is used.
We start with the standard two-dimensional advection diffusion equation for conservative scalar $C(x,y,t)$, assuming an isotropic molecular diffusion coefficient $D$, that is independent of $C$
\begin{equation}
\label{advDisp}
   \frac{\partial C(x,y,t)}{\partial t}+u(y) \frac{\partial C(x,y,t)}{\partial x}= D \frac{\partial^2C(x,y,t)}{\partial x^2}+  D \frac{\partial^2C(x,y,t)}{\partial y^2}   \: .
\end{equation}
We apply Reynolds decomposition to separate $C(x,y,t)$ and $u(y)$ into their respective means $\bar{C}(x,t)$ and $\bar{u}$, along with fluctuations $C'(x,y,t)$ and $u'(y)$. Substituting this decomposition into (\ref{advDisp}), averaging the equation, and then subtracting the average equation from (\ref{advDisp}), we obtain the equation for fluctuations
\begin{equation}
\label{flucEq}
   \frac{\partial C'}{\partial t}+(u(y)-\bar{u}) \frac{\partial \bar{C}}{\partial x} +u(y)\frac{\partial C'}{\partial x} - \frac{\partial(\overline{u'(y)C'})}{\partial x}= D \frac{\partial^2C'}{\partial x^2}+  D \frac{\partial^2C'}{\partial y^2} \: .
\end{equation}
At asymptotic times, when $t \gg \frac{(2h)^2}{D_m}$, we use the assumptions of \cite{taylor1953dispersion} (a) Gradients of fluctuations in transverse direction are larger than in longitudinal direction (b) Horizontal gradients in the mean are larger than horizontal gradients in fluctuations (c) Fluctuations evolve slowly in time. Thus, (\ref{flucEq}) reduces to
\begin{equation}
\label{flucEq2}
\frac{\bar{u}}{2D}(1-3\frac{y^2}{h^2})\frac{\partial \bar{C}}{\partial x}=\frac{\partial^2C'}{\partial y^2} \: .
\end{equation}
By integrating Equation (\ref{flucEq2}) with respect to \( y \) twice and applying no-flux boundary conditions, we obtain an expression for concentration fluctuations:
\begin{equation}
\label{Cfluc}
C'(x,y,t)=\frac{u}{8D}(2y^2-\frac{y^4}{h^2}-\frac{7}{15}h^2)\frac{\partial \bar{C}}{\partial x} \: .
\end{equation}
The mean concentration, $\bar{C}(x,t)$, is given by and can be approximated using a Taylor series expansion as follows:
\begin{equation}
\label{longC}
\bar{C}(x,t) = \frac{1}{2} \operatorname{erfc}\left(\frac{x-\bar{u}t}{\sqrt{4D_lt}}\right) \simeq \frac{1}{2} - \frac{x-\bar{u}t}{\sqrt{4\pi D_l t}} \: . %+ \mathcal{O}\left(\left(\frac{x-ut}{3\sqrt{4\pi D_l t}}\right)^3\right)
\end{equation}
Thus, at the mixing isoline defined by \(\bar{C} + C' = 0.5\),
\begin{equation}
\label{mixIso}
\frac{1}{2} - \frac{x - \bar{u}t}{\sqrt{4 \pi D_l t}} + \frac{\bar{u}}{8D} \left(2y^2 - \frac{y^4}{h^2} - \frac{7}{15} h^2 \right) \frac{1}{\sqrt{4 \pi D_l t}} \left(1 - \frac{(x - \bar{u}t)^2}{4D_l t}\right) = \frac{1}{2},
\end{equation}
we can simplify this equation and derive an explicit analytical expression for the slope of the 0.5 isoline with respect to $y$:
\begin{equation}
\label{deriv}
\frac{\partial x}{\partial y}=- \frac{\bar{u}}{2D}(y-\frac{y^3}{h^2}) \: .
\end{equation}
Also, by integrating (\ref{flucEq2}) with respect to $y$ once and applying no-flux boundary conditions, we show that the mixing interface slope $m(y)$ at asymptotic times, can be expressed as the ratio of transverse gradient of concentration fluctuations to the longitudinal gradient of mean concentration as:
\begin{equation}
\label{equality}
m(y)=\frac{\partial x}{\partial y}=- \frac{\frac{\partial C'}{\partial y}}{\frac{\partial \bar{C}}{\partial x}}=\frac{\bar{u}}{2D}(y-\frac{y^3}{h^2}) \: .
\end{equation}
That is, the slope of the 0.5 isoline $\frac{\partial x}{\partial y}$ equals the ratio of the transverse gradient of concentration fluctuations to the longitudinal gradient of mean concentration at $x = \mu_1(t)$. Lastly, due to the symmetry of the concentration field around the central line ($y = 0$), the mixing interface plateau length ($L_\infty$) can be expressed as: 
\begin{equation}
\label{IsoLeng}
L_\infty=2\int_0^h\sqrt{1+\frac{\bar{u}^2}{4D^2}(y-\frac{y^3}{h^2})^2} \, dy \: .
\end{equation}

\section*{Appendix B: Alternative Solution to Gradients at $x=\mu_1(t)$}
\renewcommand{\theequation}{B.\arabic{equation}} 
\setcounter{equation}{0} % Reset equation counter
In this appendix, we propose an alternative solution to quantify the mixing interface line mean slope $\bar{m}(t)$ by separately quantifying (1) $\overline{\nabla_y C'}(t)$, and (2) $\nabla_x \bar{C}(t)$, whose ratio is  $\bar{m}(t)$ 
\begin{equation}
\label{alter}
\bar{m}(t)=-\left[ \frac{\overline{\nabla_y C'}(t)}{\nabla_x \bar{C}(t)} \right]_{\mu_1(t)} \: .
\end{equation}
(1) Assuming that the concentration profile along any transverse level/location ($y_t$) follows a complementary error function with its center moving at an apparent velocity $u_a(t)|_{y_t}$
\begin{equation}
\label{C_erfc}
 C(x,t)_{y_t}=\frac{1}{2}erfc\left[ \frac{x-u_a(t)|_{y_t}t}{\sqrt{4D_lt}}  \right] \: .
\end{equation}
where $D_l(t)$ is the longitudinal dispersion coefficient. Hence, the concentration fluctuations at $x=\mu_1(t)$ and $y=y_t$ can be expressed as  
\begin{equation}
\label{Cfluc_erfc}
C'(t)_{y_t, \mu_1(t)}=\frac{1}{2}erfc\left[ \frac{\mu_1(t)-u_a(t)|_{y_t}t}{\sqrt{4D_lt}}  \right] -0.5 = -\frac{1}{2}erf\left[ \frac{\bar{u}-u_a(t)|_{y_t}}{2} \sqrt{\frac{t}{D_l}} \right] \: .
\end{equation}
$u_a(t)|_{y_t}$ is quantified using the history of sampled velocities due to molecular diffusion across $y \in [-h_C \:\: h_C]$ of a pulse initial condition at $y=y_t$. For a comprehensive discussion, we refer the reader to \cite{farhat2024evolution}. This approach assumes a Darcy-scale longitudinal velocity profile that remains constant in $x$. The position history distribution $P(y,t)|_{y_t}$ can be obtained by integrating the Green’s function over time and assuming a Fickian transverse diffusion. This reads:
\begin{equation}
\label{PDF}
P(y,t)|_{y_t}=\sum_{\psi=-\infty}^{\infty} \frac{|y-y_t-2\psi h_C|}{4\sqrt{\pi}Dt}\pmb{\Gamma}\left[-0.5,\frac{(y-y_t-2\psi h_C)^2}{4Dt}\right] \: ,
\end{equation}
where $\Gamma[\psi,z]$ is the upper incomplete gamma function. While (\refeq{PDF}) includes summation over an infinite range of $\psi$ virtual sources, in practical applications, $\psi$ is chosen to be large enough to ensure that the integral of $P(y,t)|_{y_t}$ over $y \in [-h_C,h_C]$ equals one. Lastly, the apparent velocity is quantified as the convolution of (a) $P(y,t)|_{y_t}$ and (b) the
mean longitudinal velocity $u(y)$ as:
\begin{equation}
\label{apparentVel}
u_a(t)|_{y_t}= \int_{-h_C}^{h_C} P(y,t)|_{y_t} u(y) \partial y   \: .
\end{equation}
Hence, the mean transverse gradient can be quantified as:
\begin{equation}
\label{TransvGrad}
\overline{\nabla_y C'}(t)_{\mu_1(t)}=\left[ \frac{C'(t,y_t=0)-C'(t,y_t=h_C)}{h_C} \right]_{\mu_1(t)} \: .
\end{equation}

(2) The gradient of the mean concentration in the longitudinal direction $\nabla_x \bar{C}(t)$ at $\mu_1(t)$ can be quantified by the maximum gradient of the one-dimensional \cite{ogata1961solution} solution for the advection diffusion equation with a continuous injection and a Heaviside initial condition. 
\begin{equation}
\label{LongGrad}
\nabla_x \bar{C}(t)= \frac{1}{2\sqrt{\pi D_l(t) t}}  \: .
\end{equation}
For transport in Poiseuille flow, one could get a closed analytical solution by replacing the longitudinal dispersion coefficient in equations (\ref{C_erfc}), (\ref{Cfluc_erfc}), and (\ref{LongGrad}) with the effective dispersion coefficient proposed in Appendix C. 
For porous media, the longitudinal dispersion coefficient $D_l(t)$ could be measured using the temporal change of the plume spreading $\mu_2(t)$ as 
\begin{equation}
\label{LongDisp}
D_l(t)=\frac{1}{2}\frac{\partial \mu_2(t)}{\partial t}  \: .
\end{equation}

\section*{Appendix C: Equality Between the Mixing Interface Length $L(t)$ and the Integral of Pore-Scale Concentration Gradients $G(t)$}
\renewcommand{\theequation}{D.\arabic{equation}} 
\setcounter{equation}{0} % Reset equation counter

In this appendix, we demonstrate that the Poiseuille-based model predicts an equality between the mixing interface length $L(t)$ and the integral of pore-scale concentration gradients $G(t)$. Specifically, we show that:
\begin{equation}
\label{equality}
L(t) = G(t) = \iint_{\Omega} \lVert \nabla C(x,y,t) \rVert \, \partial A  \: ,
\end{equation}
where $C(x, y, t) $is the concentration field, and \( \Omega \) represents the spatial domain.
To proceed, we express the magnitude of the concentration gradient in terms of its $x$- and $y$-components. This allows us to rewrite $G(t)$ in terms of the slope $m(x, y, t)$ as:
\begin{equation}
\label{Gtcom}
G(t) = \int_{-h_c}^{h_c} \int_{-\infty}^{\infty} \sqrt{\left( \nabla_y C \right)^2 + \left( \nabla_x C \right)^2} \, \partial x \, \partial y = \int_{-h_c}^{h_c} \int_{-\infty}^{\infty} \lvert \nabla_x C \rvert \sqrt{m^2 + 1} \, \partial x \, \partial y  \: ,
\end{equation}
where $m(x, y, t)$ is defined as the slope of the concentration isolines. To obtain an expression for $m(x, y, t)$, we refer to the approach detailed in Section 4.2. Solving the governing equation for the concentration fluctuation, we find that the leading-order approximation of $C'(x, y, t)$ is:
\begin{equation}
\label{Cprimexyt}
C'(x, y, t) = -\frac{6 \bar{u} h_c^2}{\pi^4 D} \cos\left(\frac{\pi y}{h_c}\right) \left(1 - e^{-\frac{\pi^2 D t}{h_c^2}}\right) \frac{\partial \bar{C}}{\partial x}  \: .
\end{equation}
Taking the derivative of $C'(x, y, t)$ with respect to $y$, we obtain an expression for $m(x, y, t)$:
\begin{equation}
\label{mxyt}
m(y, t) = \frac{\frac{\partial C'(x, y, t)}{\partial y}}{\frac{\partial \bar{C}}{\partial x}} = \frac{6 \bar{u} h_c}{\pi^3 D} \sin\left(\frac{\pi y}{h_c}\right) \left(1 - e^{-\frac{\pi^2 D t}{h_c^2}}\right)  \: .
\end{equation}
This shows that $m(y, t)$ is independent of $x$, meaning that all concentration isolines in the domain have the same slope function (constant-length isocontours).
Using the expression for $m(y, t)$, we can now approximate $G(t)$ in equation \eqref{Gtcom} as:
\begin{equation}
\label{Gtcom2}
G(t) \approx 2 \int_{0}^{h_c} \int_{-\infty}^{\infty} \lvert \nabla_x C \rvert \, m(y, t) \, \partial x \, \partial y \approx 2 \int_{0}^{h_c} \int_{-\infty}^{\infty} \frac{e^{-\frac{(x - \bar{u} t)^2}{4 D t}}}{\sqrt{4 \pi D t}} m(y, t) \, \partial x \, \partial y  \: .
\end{equation}
Solving this integral yields an explicit expression for $G(t)$:
\begin{equation}
\label{Gtexplicity}
G(t) \approx \frac{24 \bar{u} h_c^2}{\pi^4 D} \left(1 - e^{-\frac{\pi^2 D t}{h_c^2}}\right) \: .
\end{equation}
Finally, approximating the expression for $L(t)$ (as outlined in equation \eqref{LengthAprox}) and using the leading-order approximation from equation \eqref{b1_2}, we find:
\begin{equation}
\label{LequalG}
L(t) \approx 2 h_c \bar{m}(t) \approx \frac{24 \bar{u} h_c^2}{\pi^4 D} \left(1 - e^{-\frac{\pi^2 D t}{h_c^2}}\right)  \: .
\end{equation}
Thus, we conclude that  $L(t) = G(t)$, demonstrating that the proposed Poiseuille-based model for $L(t)$ is consistent with the equality observed in Figure (\ref{fig:mixing interface}) for a porous medium.
\section*{Appendix D: Quantifying the Effective Dispersion Coefficient $D_e(t)$ Using the Mixing Interface Mean Slope $\bar{m}(t)$}
\renewcommand{\theequation}{D.\arabic{equation}} 
\setcounter{equation}{0} % Reset equation counter
In this appendix, we demonstrate how to directly calculate the effective dispersion coefficient $D_e(t)$ for a line source using the mixing interface mean slope, $\bar{m}(t)$. To begin, we introduce the following dimensionless quantities:
\begin{equation}
\label{dimensionless}
 \hat{x}=\frac{x}{h}, \quad \hat{y}=\frac{y}{h}, \quad \hat{C}=\frac{C}{C_{ref}}, \quad \hat{t}= \frac{t}{t_D}, \quad \hat{u}=\frac{\bar{u}h}{D},
\end{equation}
Using Reynolds decomposition, as discussed in Appendix A, the governing equation for the average concentration can be expressed as:
\begin{equation}
\label{average}
\frac{\partial{\hat{\bar{C}}}}{\partial{\hat{t}}} + Pe \frac{\partial{\hat{\bar{C}}}}{\partial{\hat{x}}} = \frac{\partial^2{\hat{\bar{C}}}}{\partial{\hat{x}}^2} - \frac{\partial{\left(\overline{\hat{u}'(\hat{y}) \hat{C'}} \right)}}{\partial{\hat{x}}} \: .
\end{equation}
Using the closure approximation from equation (\ref{DiogoCfluc}), \cite{bolster2011mixing} showed that the average model can be closed as follows:
\begin{equation}
\label{average}
\frac{\partial{\hat{\bar{C}}}}{\partial{\hat{t}}} + Pe \frac{\partial{\hat{\bar{C}}}}{\partial{\hat{x}}} = \hat{D}_a(t) \frac{\partial^2{\hat{\bar{C}}}}{\partial{\hat{x}}^2}+\phi(\hat{x},\hat{t}) \: ,
\end{equation}
where $\phi(\hat{x},\hat{t})$ is a memory function accounting for the initial condition's effects, and $\hat{D}_a(t) = \frac{D_a(t)}{D}$ is the normalized, time-dependent, apparent Taylor-dispersion coefficient:
\begin{equation}
\label{dispersion}
D_a(t) = D - D \int_0^1{\hat{u}'(\hat{y}) b_1(\hat{y}, \hat{t})} \, \mathrm{d} \hat{y}  \: .
\end{equation}
An analogous approach to that in Section 4.2 yields:
\begin{equation}
\label{b1nonDim}
b_1(\hat{y}, \hat{t}) = \sum_{\alpha=1}^{\infty} \frac{12 Pe \left( 1 + (-1)^{\alpha} \right) \left( 1 - e^{-\alpha^2 \pi^2 \hat{t}} \right) \cos(\alpha \pi \hat{y})}{\alpha^4 \pi^4}  \: .
\end{equation}
Approximating \eqref{b1nonDim} using the leading-order term ($\alpha=2$), we solve for the apparent dispersion coefficient:
\begin{equation}
\label{dispersion2}
D_a(t) = D + D \Bigg[ \frac{9}{2 \pi^6} Pe^2\left( 1 - e^{-4 \pi^2 \hat{t}} \right) \Bigg]  \: .
\end{equation}
\cite{dentz2007mixing} showed that the global effective dispersion coefficient for a line source can be calculated using the apparent dispersion as:
\begin{equation}
\label{effective}
D_e(t)=2D_a(t)-D_a(2t) \: .
\end{equation}
Substituting \eqref{dispersion2} into \eqref{effective} and rearranging, we obtain the explicit solution for the effective dispersion coefficient:
\begin{equation}
\label{effectiveExplict}
D_e(t)=D \Bigg[1+\frac{9}{2\pi^6}Pe^2\left( 1 - e^{-4 \pi^2 \hat{t}} \right)^2\Bigg] .
\end{equation}
Finally, by recalling the expression for the transient mixing interface mean slope, as given in equation \eqref{b1_2}, we show that the global effective dispersion coefficient can be directly quantified using the mean slope $\bar{m}(t)$ as:
\begin{equation}
\label{effecSlope}
D_e(t)=D(1+1.25\bar{m}^2(t)) .
\end{equation}
\end{document}